\documentclass[% 
reprint, 
amsmath,amssymb, aps, superscriptaddress,prb,nofootinbib
]{revtex4-2}
\usepackage[dvipsnames]{xcolor}
\usepackage{tikz}
\usepackage{siunitx}
\usepackage{amsmath}
\usepackage{amssymb}
\usepackage{mathrsfs}
\usepackage[percent]{overpic}
\usepackage{hyperref}
\usepackage{bm}
\usepackage[capitalise]{cleveref}
\usepackage{comment}
\usepackage{nicefrac}
\usepackage[percent]{overpic}
\usepackage{array, makecell}
\usepackage{mathtools}
\usepackage{amsmath}
\usepackage{graphicx}% Include figure files
\usepackage{dcolumn}% Align table columns on decimal point
\usepackage{bm}% bold math

\usepackage[normalem]{ulem} % temporarily - \sout command
\newcommand{\vampire}{\textsc{vampire} }

\begin{document}

\title{Magnon dispersion and spin transport in CrCl$_3$ bilayers under different strain-induced magnetic states}

\author{Verena Brehm}
 \email{verena.j.brehm@ntnu.no}
\affiliation{Center for Quantum Spintronics, Department of Physics, Norwegian University of Science and Technology, Trondheim, Norway}%
\author{Stefan Stagraczy\'nski}
\affiliation{Faculty of Physics, ISQI, Adam Mickiewicz University in Pozna\'n, ul. Uniwersytetu Pozna\'nskiego 2, 61-614 Pozna\'n, Poland}
\author{J\'ozef Barna\'s}
\affiliation{Faculty of Physics, ISQI, Adam Mickiewicz University in Pozna\'n, ul. Uniwersytetu Pozna\'nskiego 2, 61-614 Pozna\'n, Poland}
\author{Anna Dyrda{\l}}
\affiliation{Faculty of Physics, ISQI, Adam Mickiewicz University in Pozna\'n, ul. Uniwersytetu Pozna\'nskiego 2, 61-614 Pozna\'n, Poland}
\author{Alireza Qaiumzadeh}\affiliation{Center for Quantum Spintronics, Department of Physics, Norwegian University of Science and Technology, Trondheim, Norway}%

\begin{abstract}
Atomically-thin van der Waals magnetic materials offer exceptional opportunities to mechanically and electrically manipulate magnetic states and spin textures. The possibility of efficient spin transport in these materials makes them promising for the development of novel nanospintronics technology.
Using atomistic spin dynamics simulations, we investigate magnetic ground state, magnon dispersion, critical temperature, and magnon spin transport in CrCl$_3$ bilayers in the absence and presence of compressive and tensile strains.  
We show that in the presence of mechanical strain, the magnon band gap at the $\Gamma$ point and the critical temperature of the bilayer are increased. Furthermore, our simulations show that the magnon diffusion length is reduced in the presence of strain. 
Moreover, by exciting magnons through the spin Seebeck effect and spin Hall-induced torque, we illustrate distinctions between magnon spin transport in the antiferromagnetic state, under compressive strains, and ferromagnetic states, under tensile strains or in the unstrained case.
\end{abstract}

\maketitle

%----------------------------------------------
\section{Introduction}
%----------------------------------------------
Two-dimensional (2D) magnetic systems \cite{RevModPhys.92.021003,Mounet2018,Gibertini2019,doi:10.1126/science.aav4450} 
represent a novel and promising platform for the next generation of magnonic \cite{Chumak2015,Cornelissen2015LongDistSpinTransp,MagSpinTransportChemPot} and spintronic \cite{Kaverzin_2022,2DspintronicsReview,MagneticGenom2DvdW2022}  nanodevices. This is attributed to their highly tunable magnetic and electronic properties, making them an ideal testbed for probing novel exotic phenomena \cite{Bedoya_Pinto_2021,Rodin_2020,Song2021,Gao2021,PhysRevLett.127.017701,generalCrCladvantages,kløgetvedt2023tunable}.
With the demonstration of long-range magnetic order in monolayer and few-layer thickness van der Waals (vdW) materials
\cite{doi:10.1126/science.aav4450,Magnetism_in_2D_van_der_Waals_materials,RamanFePs3Wang2016,Bedoya_Pinto_2021,Huang2017}, the class of chromium trihalides (CrX$_3$, X=Cl, Br, I) with honeycomb lattice structure has come into focus. It has been shown that in contrast to CrI$_3$ that shows a topological band gap in the magnon dispersion at the Dirac points \cite{ChenPRX2018,ChenSecondExpDispersion_KitaevAndDMI,ChenPRX2021,brehm2023topological}, CrCl$_3$  hosts massless Dirac magnons at the K and K' points \cite{Chen_2022,Schneeloch2022}.

In particular, the possibility of tuning the magnonic properties in these layered magnetic vdW materials through external stimuli, such as electrostatic gating, magnetic fields, and mechanical strains \cite{menichetti2023electrical,Edstrom2022,Basak_2023,Ren2023,DMIandKitaevinCrI3analytical,PhysRevB.105.L100402,PhysRevB.101.125111,PhysRevB.106.125103,PhysRevLett.123.237207,soenenTunableMagTop,PhysRevMaterials.4.094004,generalCrCladvantages,AliDFT}, is intriguing for application in novel nanospintronic technology. 
Recent theoretical studies have demonstrated controllable manipulation of the magnetic ground state in the CrCl$_3$ mono- and bilayer through mechanical strains \cite{generalCrCladvantages,AliDFT}. Biaxial strains have the potential to induce transitions between ferromagnetic (FM) and antiferromagnetic (AFM) phases. Additionally, they can alter the magnetic ground state from uniaxial out-of-plane (OOP) to biaxial easy-plane (EP) magnetic states. 

Although there are several theoretical and experimental studies on the static magnetic behavior of these materials, only a few recent experiments have explored magnon spin transport in atomically-thin vdW FM and AFM materials via either thermal \cite{PhysRevB.107.094428,PhysRevB.106.224409,PhysRevB.101.205407,PhysRevX.9.011026} or electrical \cite{LongDistMagTransportWees} mechanisms. To demonstrate the potential of vdW magnets for spintronic and magnonic nanotechnology, additional theoretical and experimental studies under various external stimuli are essential.

In order to make a theoretical ground for further studies in magnon dynamics in vdW materials, in this paper, we investigate magnon transport in a bilayer of CrCl$_3$ under biaxial compressive and tensile strains. Using atomistic spin dynamics simulations \cite{vampire}, we compute magnon dispersion and magnon transport, where the latter is excited through the thermal spin Seebeck effect (SSE) \cite{Uchida_2010,PhysRevX.6.031012,PhysRevB.93.014425,SSE,Adachi_2013,Berakdar2013,Berakdar2015,UlrikeSEEtransportsimulation, UlrikeSSE2,MookSEEandNernstEff,MarkusSkyrHallEff} and the electrical (anomalous) spin-Hall torque (SHT) mechanism \cite{lebrun_long-distance_2018,lebrun_long-distance_2020,Hirsch1999,Zhang,AFMtransportPRBarticle,InverseAnomalousSHeffect}. Three sets of effective spin parameters represent three magnetic states that can be achieved through mechanical strains \cite{AliDFT}: an EP AFM state in the presence of a compressive strain (negative strain), an EP FM state at the absence of strain, and an OOP FM state in the presence of a tensile strain (positive strain). It is worthwhile to mention that while we use CrCl$_3$ spin interaction parameters in this study, the qualitative findings of this article can be applied to a wide class of vdW magnetic layers.

The rest of the paper is structured as follows: First, we present our system setup, its effective spin model, and our spin dynamic simulation method in \cref{sec:model}. In \cref{sec:dispersions}, we compute the magnon dispersion and critical temperature in the CrCl$_3$ bilayer under various mechanical strains. In \cref{sec:transport}, we investigate magnon spin transport and compute the magnon diffusion length in different magnetic states of the system. Finally, we discuss and conclude the findings and their practical implications in \cref{sec:conclusion}.

%----------------------------------------------
\section{Model and method} 
\label{sec:model}
%----------------------------------------------
This section presents our system geometry, effective spin Hamiltonian, and the atomistic spin dynamic simulation method employed to calculate magnon dispersion and magnon spin transport.

\begin{figure}
        \centering
       \includegraphics[width=\columnwidth]{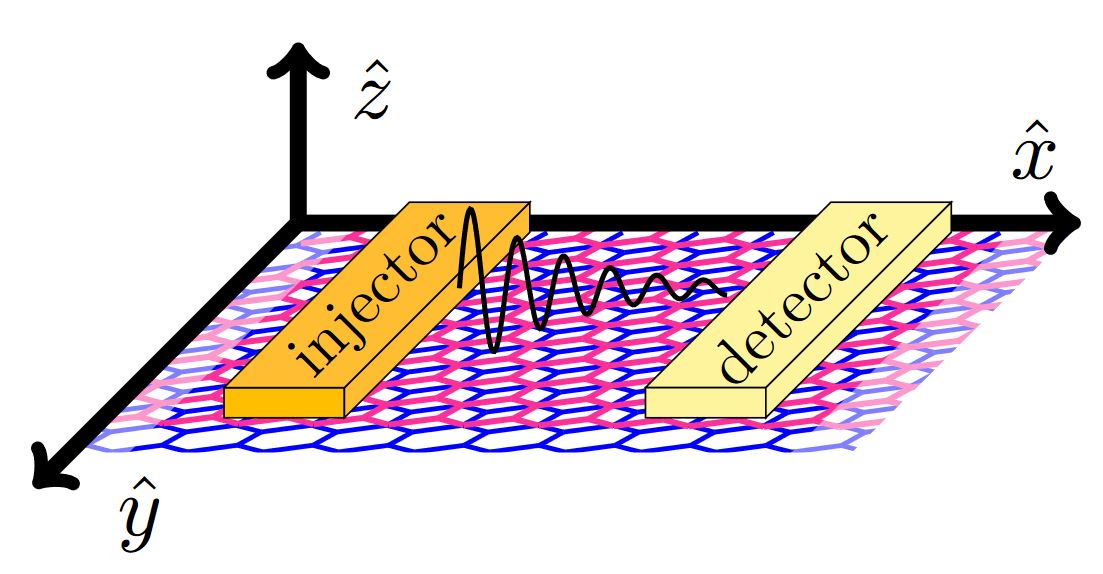}
        \caption{Schematic setup for magnon spin transport simulation in a  CrCl$_3$ bilayer with honeycomb lattice structure: magnons, denoted by a solid black damped wave, are injected by the left lead  through the SSE and/or SHT mechanisms, and propagate via a diffusive transport to the right lead along the $x$ direction, where they are detected through the inverse (anomalous) spin Hall effect.} 
        \label{fig:setupTransport}
    \end{figure}
%----------------------------------------------
\subsection{Device geometry and setup}
%----------------------------------------------
To investigate spin transport in a CrCl$_{3}$ bilayer, we use a setup depicted in \cref{fig:setupTransport}.
It emulates experimental systems where spin-transport measurements are performed in the so-called nonlocal geometry \cite{lebrun_long-distance_2020,LongDistMagTransportWees,AFMtransportPRBarticle}. In this geometry, magnons are injected from the left lead via either SSE \cite{Uchida_2010,PhysRevX.6.031012,lebrun_long-distance_2018} and/or the (anomalous) SHT \cite{InverseAnomalousSHeffect,SpinHallEffects,lebrun_long-distance_2018} and propagate through the system along the $x$ direction until the magnon spin current signal is detected on the detector lead.
In magnon spin transport experiments, the injector and detector can be either a heavy metal or an FM metal with strong spin-orbit coupling.
In the first case, the (inverse) spin Hall effect only allows for the injection (detection) of in-plane spin signals \cite{SpinHallEffects}. 
However, in the second case, the (inverse) anomalous spin Hall effect enables the injection (detection) of OOP spin signals \cite{InverseAnomalousSHeffect}. We will address spin injection and detection processes in more detail in the next sections.

    \begin{figure}
        \centering
        \includegraphics[width=\columnwidth]{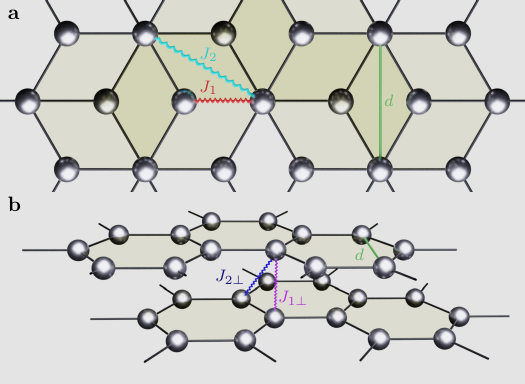}
        \caption{Atomic model depicting magnetic Cr atoms arranged in a honeycomb lattice within a CrCl$_3$ bilayer with rhombohedral stacking. Figures present a top view (a) and a perspective view (b). The red and cyan wavy lines in (a) illustrate the \textit{intra}layer NN, $J_1$, and NNN, $J_2$, respectively. The purple and blue wavy lines in (b) denote \textit{inter}layer NN, $J_{1\perp}$, and NNN, $J_{2\perp}$, coupling, respectively. $d\approx\SI{0.6}{nm}$ is the lattice constant.}
        \label{fig:Rhstacking}
    \end{figure}
%----------------------------------------------
\subsection{Effective spin Hamiltonian of CrCl$_3$ bilayer}
%----------------------------------------------
A CrCl$_3$ bilayer consists of two atomic planes of magnetic Cr atoms surrounded by nonmagnetic Cl atoms. Within each atomic plane, the magnetic Cr atoms are arranged on a honeycomb lattice, and the two planes are stacked rhombohedrally. The system thus has four spins per magnetic unit cell. The effective spin interaction between magnetic Cr atoms can be modeled by the following Hamiltonian \cite{AliDFT,doi:10.1021/acs.nanolett.9b01317}:
\begin{align} 
\mathscr{H} =& - \sum_{i<j} J_{ij} \mathbf{S}_i \cdot \mathbf{S}_j - K_x \sum_i \left(\mathbf{S}_i \cdot \hat{e}_x\right)^2 \nonumber\\ &- K_z \sum_i \left(\mathbf{S}_i \cdot \hat{e}_z \right)^2.
\label{eq:hamiltonian}
\end{align}
In this spin Hamiltonian, $\mathbf{S}_i$ is a unit vector that carries the direction of the atomic magnetic moment at site $i$, $J_{ij}$ denotes the symmetric Heisenberg exchange coupling between sites $i$ and $j$. This exchange term includes the \textit{intra}layer nearest-neighbor (NN) $J_{1}$ and next-nearest neighbor (NNN) $J_2$ coupling, as well as \textit{inter}layer NN $J_{1\perp}$, and NNN $J_{2\perp}$ coupling, see \cref{fig:Rhstacking}. We also introduce two single-ion magnetic anisotropy axes along the in-plane $x$ and OOP $z$ directions, parameterized by $K_x$ and $K_z$, respectively. 
Since the Dzyaloshinskii–Moriya interactions are very weak in this system, we neglect them in our effective spin Hamiltonian \cite{AliDFT,ESTERASstrainDMI,WebsterstrainDMI}.

Mechanical strains can modify both the sign and strength of Heisenberg exchange coupling $J_{ij}$ and magnetic anisotropy $K_{x (z)}$ \cite{AliDFT}. 
The exchange coupling parameters $J_{ij}$ are modified by strain, leading to an AFM order at -5\% strain, and an FM order at 0\% and +5\% strain \cite{AliDFT}.
Furthermore, for both the unstrained and the -5\% compressive strained case, the magnetic ground state lies along the $x$ direction inside the easy $x$-$y$ plane, with $K_x>0$ and $K_z<0$, which leads to an EP magnetic state. For the +5\% tensile strain case, the magnetic ground state lies along the $z$ direction, $K_z>0$, resulting in an OOP FM state.  
The parameters used in our spin dynamics simulations are listed in \cref{tab:anisos}. In order to reduce computational cost, we do not include dipolar interactions in our simulations. This can be justified by the fact that in 2D bulk vdW magnetic materials, dipolar interactions are mainly reduced to an effective magnetic anisotropy that slightly shifts the magnon spectrum \cite{Hussain_2022,2018JPCM...30y5803B}. This has no qualitative impact on our findings, since it only leads to a minor modification of the magnetic anisotropy constants.

%-------------------------------- TAB I --------------------------------
\begin{table}[t]
\centering
\caption{The value of spin interaction parameters under various strains \cite{AliDFT}, introduced in the spin Hamiltonian \cref{eq:hamiltonian}: NN and NNN intra-layer, $J$, and inter-layer, $J_\perp$, Heisenberg exchange parameters, and magnetic anisotropy $K_{x(z)}$.
The magnetic ground state is sketched using the sublattice spins.}
\scalebox{0.96}{\begin{tabular}{cccc} \hline
\textbf{Parameter} & \textbf{0\% strain} & \textbf{+5\% strain} & \textbf{-5\% strain} \\ \hline
$J_1$ & \SI{1.53}{meV}  & \SI{2.39}{meV} & -\SI{2.94}{meV}\\
$J_2$ & \SI{0.29}{meV}  & \SI{0.24}{meV} & \SI{0.44}{meV} \\
$J_{1\perp}$ & \SI{0.07}{meV}  & \SI{0.11}{meV} & \SI{0.01}{meV} \\
$J_{2\perp}$ & \SI{0.07}{meV} & \SI{0.08}{meV}  & \SI{0.07}{meV} \\
$K_x$ & $\SI{0.06}{meV} $ & $0$  & $\SI{0.06}{meV} $  \\
$K_z$ & $-\SI{0.5}{meV}$   & $\SI{1.25}{meV}$  & $\SI{-4.75}{meV}$\\ \hline
\textbf{\small{Magnetic state}} & $\substack{\rightarrow\\[-1em] \rightarrow} $  & $\uparrow \uparrow $ & $\substack{\rightarrow\\[-1em] \leftarrow} $\\
&  EP FM  & OOP FM &  EP AFM\\
\hline
\end{tabular}}
\label{tab:anisos}
\end{table}

%----------------------------------------------
\subsection{Atomistic spin dynamics simulations}
\label{sec:AtomSpinSim}
%----------------------------------------------
Spin dynamics at finite temperature is governed by the stochastic Landau–Lifshitz-Gilbert (sLLG) equation \cite{vampire}: 
\begin{equation}
\label{eq:LLG}
\frac{\partial \mathbf{S}_i}{\partial t} = -\frac{\gamma}{1 + \alpha_0^2} \left[\mathbf{S}_i \times \mathbf{B}^\mathrm{eff}_i + \alpha_0 \mathbf{S}_i\times (\mathbf{S}_i \times  \mathbf{B}^\mathrm{eff}_i )\right],
\end{equation}
where $\gamma$ is the gyromagnetic ratio, $\alpha_0$ denotes the Gilbert damping parameter, and $\mathbf{B}^\mathrm{eff}_i$ is the effective magnetic field at site $i$.
The effective field $\mathbf{B}^\mathrm{eff}_i$ consists of two terms: a deterministic term, related to the spin Hamiltonian, and a stochastic term, related to the thermal fluctuations, 
\begin{equation}
    \mathbf{B}^\mathrm{eff}_i = -\frac{1}{\mu_s} \frac{\partial \mathscr{H}}{\partial \bm{S}_i} + \bm{\xi}^{\rm (th)}_i, \label{EffectiveField}
\end{equation}
where $\mu_s$ is the atomic spin moment. The stochastic term at the low-frequency regime
is modeled as Gaussian thermal noise,
\begin{subequations}
\begin{align}
    & \langle\bm{\xi}^{\rm (th)}_i(t)\rangle =0, \\
    & \langle\xi_{i,m}^{\rm (th)}(t) \xi_{j,n}^{\rm (th)}(t')\rangle = \frac{2 \alpha_0 k_\mathrm{B}T }{\gamma \mu_s} \delta_{ij}\delta_{mn}\delta(t - t'),
\end{align}
\end{subequations}
where $k_\mathrm{B}$ is the Boltzmann constant and $m,n = \{ x,y,z \}$ denote spatial components. 

To investigate the spin dynamics within our system, we utilize a stochastic Heun algorithm implemented in the \vampire code \cite{vampire, vampireurl} for the numerical solution of the sLLG equation.

%----------------------------------------------
\subsection{Spin injection and spin detection} \label{subsecTransportmodelling}
%----------------------------------------------
\emph{(i) Detection of spin signal via spin pumping --} 
The spin signal at the distance $x$ from the injector is determined by computing the local spin accumulation at the detector interface \cite{ArneSpinPumping, PhysRevB.102.020408,tang2023absenceCrossSublTerms},
    \begin{equation} \label{eq:spinAccum}
        \bm{\mu}(x) :=  G_r^{\uparrow \downarrow}  \sum_{i=1}^N \left<\big[\bm{S}_{i}(t)\times \dot{\bm{S}}_{i}(t)\big]\right>_t,
    \end{equation}
where the sum runs over sites at the interface between the detector and the magnetic bilayer, $G_r^{\uparrow \downarrow}$ is the real part of the spin mixing conductance \cite{ArneSpinMixingConductance}, and $\left<\cdot \right>_t$ denotes a time average that is conducted after steady state is reached. As we already mentioned, the spin accumulation at the interface is converted to a charge voltage at the lead via the inverse (anomalous) Hall effect.
In this article, we are interested in the dc spin signal, and thus, we only present the component of the spin accumulation that is parallel to the magnetic ground state.

\emph{(ii) Injection of spins via SSE and SHTs --}
The spin signal can be generated through a thermal gradient, via SSE, and/or electrically through spin Hall effects, via (anomalous) SHTs.

In the first scenario, a temperature gradient is applied to generate a spin voltage through the so-called SSE \cite{Uchida_2010,PhysRevX.6.031012,lebrun_long-distance_2018, SSE,Adachi_2013,Berakdar2013,Berakdar2015,PhysRevB.108.224420}. The increase in temperature in the injector results in a higher local occupancy of thermal magnon modes compared to the rest of the system, which maintains a lower temperature. This results in an incoherent magnon current flowing from the hotter side toward the cooler side \cite{UlrikeSEEtransportsimulation, UlrikeSSE2,MookSEEandNernstEff,MarkusSkyrHallEff}.

In the second scenario, the charge-induced (anomalous) spin Hall effect in the injector layer gives rise to interfacial spin torques \cite{lebrun_long-distance_2018,lebrun_long-distance_2020,Cornelissen2015LongDistSpinTransp}. 
These SHTs can be modeled using the following effective field \cite{Andrea_STTinVampire_2023} that is added to the effective magnetic field, \cref{EffectiveField},
\begin{align}
        \mathbf{B}^{\tau}_i &= \tau_0^{P} \left(\bm{p}-\alpha_0 \bm{S}_i\times \bm{p}\right)+ \tau_0^{R}\left(\bm{S}_i\times \bm{p}+\alpha_0 \bm{p}\right).
    \end{align} 
Here $\bm{p}$ is the spin polarization direction, induced by the (anomalous) spin Hall effect. The first term is the precession-like field, parameterized by $\tau_0^P$, and the second term is the relaxation-like field, parameterized by $\tau_0^R$. These two parameters are related to the material parameters and charge current density.

The relaxation-like torque is responsible for generating an effective magnon chemical potential \cite{Demidov_2017} at the interface. Consequently, it facilitates the flow of the magnon current through the system.
On the other hand, the precession-like torque only adjusts the eigenfrequencies of the magnon eigenstates \cite{Andrea_STTinVampire_2023}.

In our simulations, we set $\bm{p}$ parallel to the ground-state spin direction. As we discussed earlier, in experiments, an in-plane and OOP spin polarization direction $\bm{p}$ can be generated by spin Hall and anomalous spin Hall effects, respectively \cite{InverseAnomalousSHeffect}.
Note that in this scenario, where $\bm{p}$ is parallel to the magnetic ground state, finite thermal fluctuations are required to generate a finite spin torque at the interface \cite{AFMtransportPRBarticle}.   

\begin{figure*}
    \centering
\includegraphics[width=0.95\textwidth]{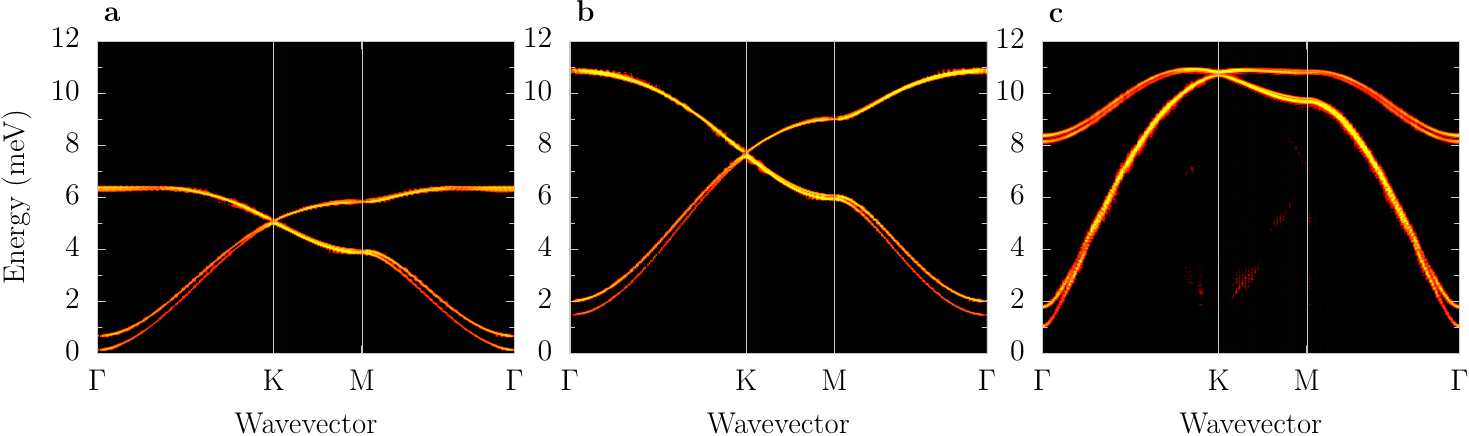}
    \caption{Magnon dispersion relation for (a) an unstrained case 0\% (biaxial EP FM state), (b) a tensile strain +5\% (uniaxial OOP FM state), and (c) a compressive strain -5\% (biaxial EP AFM state).}
    \label{fig:dispersionrelations}
\end{figure*}
%---------------------------------------------------------------------
\section{Magnon dispersion and critical temperature} 
\label{sec:dispersions}
%---------------------------------------------------------------------
To calculate the magnon dispersion through atomistic spin dynamics simulations \cite{vampire}, we set up a  CrCl$_3$ bilayer system with rhombohedral stacking. First, we determine the magnetic ground state of the system. Following this, we introduce a small random deviation around the equilibrium spin directions. Next, we let spins evolve according to the sLLG equation, \cref{eq:LLG}, and compute the temporal and spatially resolved transverse spin components. Finally, we find the dispersion curve by performing a Fourier transform in both the time and space domains.

The magnon dispersion of the CrCl$_3$ bilayer is shown in \cref{fig:dispersionrelations} for all the three magnetic states. Since the honeycomb magnetic unit cell carries two magnetic atoms, there are two magnon branches per layer, one at low frequencies, acoustic-like, and one at high frequencies, optic-like. Thus, in total, the bilayer system has four magnon branches. Magnon branches can be partially (non)degenerate in some regions of the magnetic Brillouin zone.  
The degeneracy of the two acoustic-like magnon branches at the $\Gamma$ point can be lifted by either the interlayer exchange couplings or the presence of a hard-axis magnetic anisotropy.
While the easy-axis magnetic anisotropy opens a magnon band gap at the $\Gamma$ point, the amplitude of the exchange couplings modifies the slope of the branches and also determines the magnon bandwidth.

Figure~\ref{fig:dispersionrelations}(a) shows the magnon dispersion of the unstrained case, with a biaxial EP FM state. The degeneracy of acoustic-like magnon bands around the $\Gamma$ point is lifted because of a finite hard-axis magnetic anisotropy $K_z<0$ and its lowest branch is gapped because of a finite easy-axis magnetic anisotropy $K_x>0$. Around the K point, there is a massless Dirac-like magnon dispersion with a fourfold degenerate crossing at the K point.

Figure~\ref{fig:dispersionrelations}(b) presents the magnon dispersion of the bilayer system in the presence of tensile strain with OOP FM state. In this case there is only a uniaxial easy-axis anisotropy $K_z>0$ which is larger than its counterpart in the unstrained case and thus creates a larger magnon band gap at the $\Gamma$ point. The degeneracy of acoustic-like magnon branches around the $\Gamma$ point is lifted because of interband exchange interactions. The magnon dispersion at the K-point maintains its fourfold degenerate crossing in this case as well. The magnon bandwidth is increased in this case compared to the zero strain case.

Figure~\ref{fig:dispersionrelations}(c) shows the magnon dispersion of the bilayer system in the presence of compressive strain with a biaxial EP AFM ground state. The linear magnon dispersion around the $\Gamma$ point, characteristic of AFM magnons, is different from the FM cases with a parabolic dispersion. The magnon gap at the $\Gamma$ point and lifting the degeneracy of the two acoustic-like branches are caused by the presence of an easy-axis $K_x>0$ and a hard-axis $K_z<0$ magnetic anisotropy, respectively.

The characteristics of magnon modes vary across the three distinct magnetic states engineered by mechanical strains. In FM systems, magnon eigenstates have right-handed helicity, whereas in AFM systems, both left- and right-handed helicities are allowed \cite{Introduction_to_AFM_magnons_Rezende,AFMtransportPRBarticle}. Moreover, the ellipticity of magnon eigenmodes varies across the three scenarios owing to distinct magnetic anisotropies. In the magnetic easy-axis case, the magnon eigenmodes are circularly polarized. In the easy-plane case, however, magnons are elliptically polarized, attributed to the magnetic hard-axis anisotropy \cite{Introduction_to_AFM_magnons_Rezende,AFMtransportPRBarticle}.

In our 2D magnetic model, \cref{eq:hamiltonian}, without long-range dipolar interactions, the magnon gap at the $\Gamma$ point mainly determines the critical temperature of the magnetic states \cite{Maleev1,Maleev2,PhysRevB.33.6519,PhysRevB.43.6015,PhysRevB.46.861,PhysRevLett.77.386,BreakingThroughMerminWagner}, which is dictated by the Hohenberg–Mermin–Wagner theorem \cite{PhysRevLett.17.1133,PhysRev.158.383,Coleman}. Due to variations in Heisenberg exchange couplings and magnetic anisotropies, induced by strain fields, the CrCl$_3$ bilayer shows different critical temperatures across its three magnetic states. We find the critical temperature, using specific heat calculations, for three magnetic states, see the Appendix for technical details,
\begin{subequations} \label{eq:Tc}
    \begin{align} 
        &T_{\rm{c}}^\text{0\%} \approx \SI{18}{K},\\
        &T_{\rm{c}}^\text{+5\%} \approx \SI{26}{K},\\
        &T_{\rm{c}}^\text{-5\%} \approx \SI{32}{K}.
    \end{align}
\end{subequations}

These results show the viability of 2D magnetic bilayers for novel magnonic technology. In the next section, we show how the spin angular momentum, as an information carrier, can be transferred by these magnons in the system.

%-------------------------------------------------------------------------
\section{Magnon spin Transport in bilayer ${\mathbf{CrCl_{3}}}$}
\label{sec:transport}
%-------------------------------------------------------------------------
As we have discussed in \cref{subsecTransportmodelling}, the spin voltage can be created by either SSE or SHT in a nonlocal geometry setup; see \cref{fig:setupTransport}. 
It is worth mentioning that in SHT experiments, an SSE is also generated because of the parasitic Joule heating created by an applied low-frequency charge current in the injector and contributes to the measured total spin voltage in the detector \cite{lebrun_long-distance_2020,lebrun_long-distance_2018,Cornelissen2015LongDistSpinTransp}. To discriminate between the spin voltage induced by nonthermal SHT and thermal SSE, the first- and second-harmonic voltages are measured in the detector, respectively. The even component of the spin voltage, which is quadratic in the applied charge current and related to the thermal SSE contribution, is the average of measured spin signals for two opposite charge current polarities in the injector. The odd component, which is linear in the applied charge current and describes the nonthermal SHT contribution, is computed from the difference between spin voltages generated by two opposite charge current polarities. In the following we also compute these even and odd spin signals.

%----------------- TAB II --------------% 
    \begin{table}[tbp]
	\centering
	\caption{Material parameters used for the magnon spin transport simulations.}
 \scalebox{0.92}{
	\begin{tabular}{@{}lccc@{}} \hline
		\textbf{Quantity}                 & \textbf{Symbol}            & \textbf{Value}               & \textbf{Unit}                 \\ 
        %Unit cell size & $\[a_x, a_y, a_z\]$ & $\[7, 12, 30\]$ & $\si{\nano \meter} $\\
        Length & $L_x$ & $\SI{500}{}$ & $\si{nm}$ \\
		Width & $L_y$ & $\SI{50}{}$ & $\si{nm}$ \\
        Lattice constant & $d$ & $0.6$ & $\si{nm} $\\
		%Thickness & $L_z$ & $3$ & $\si{nm} $\\
		Time step &           $\Delta t$            &      $\SI{1}{}$             &                    $\si{\femto\second}$ \\
		Bulk Gilbert damping \cite{Kapoor_2020} &           $\alpha_0$           &   $\SI{2e-3}{}$                    &                      -\\
        Gilbert damping at left (right) edge &           $\alpha_L (\alpha_R)$           &   $0.5$ ($0.9$)                    &                      -\\ \hline 
        \textbf{SHT Mechanism} & & \\
        Spin torques in FM (AFM) state & $\tau_0^P = \tau_0^R$ & $0.01 (0.2)$ & \si{T}\\
        Background temperature & $T_0$ & 1 & \si{K} \\ 
        %Injector temperature  & $T_\text{inj}=T_0$ & 1 & \si{K} \\
        \hline 
        \textbf{SSE Mechanism} & & \\
        %Spin torques & $\tau_0^P = \tau_0^R$ & 0 & \si{T}\\
         Injector temperature & $T_\text{inj}$ & 5 & \si{K} \\ 
         Background temperature & $T_0$ & 1 & \si{K} \\ 
	\end{tabular}}
	\label{tab:TransportSimParams}
    \end{table}
%----------------- TAB II (end) --------------% 

In the presence of Gilbert damping, magnons diffusely propagate through the system along the $x$ direction in our setup, \cref{fig:setupTransport}. The spatial dependence of  nonequilibrium spin accumulation, far from the injector, can be modeled by \cite{WeesMagnonDiffusion}, 
    \begin{equation} \label{eq:decay}
        \delta \bar{\mu}(x):= \frac{\mu(x)-c}{\mu_0} =  \exp\left(-\frac{x-x_0}{\lambda}\right),
    \end{equation}
where $\lambda$ is the magnon spin diffusion length, $x_0\gg d$, and $c$ is the background thermal equilibrium spin accumulation. 
The spin accumulation is normalized to $\mu_0 \equiv \mu(x_0)$. As we mentioned earlier, we only compute the component of spin accumulation that is parallel to the magnetic ground state in each case.

\begin{figure}[htbp]
        \centering
        \includegraphics[width=\linewidth]{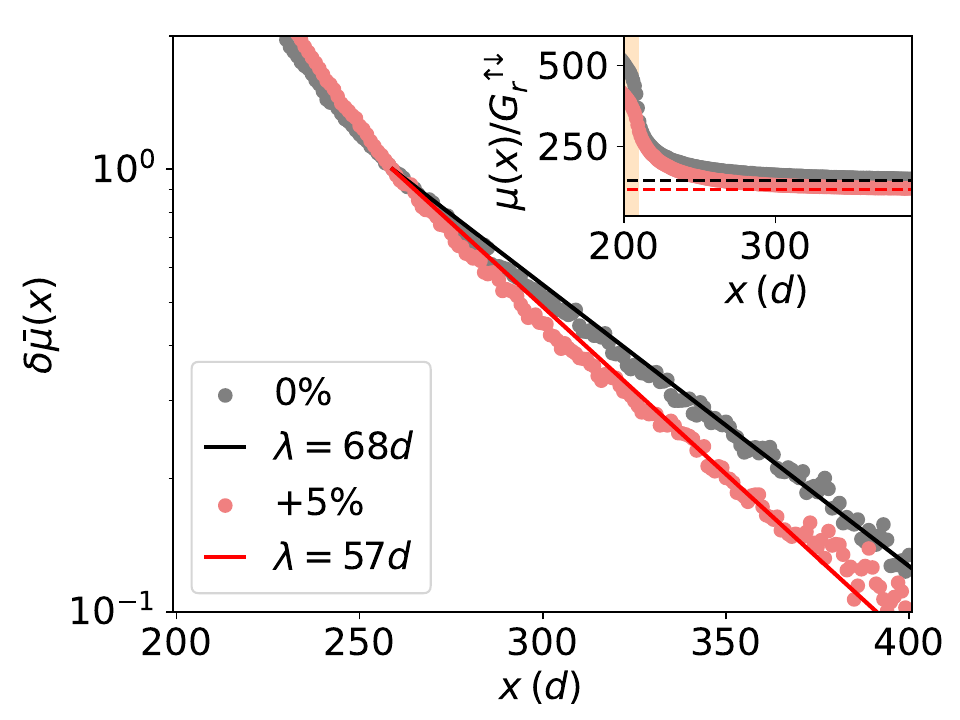}
        \caption{Spatial dependence of the thermal SEE-induced nonequilibrium spin accumulation for EP (unstrained) and OOP (+5\% strain) FM states. There is no spin signal in -5\% strain case with an AFM state in this scenario. 
        The extracted spin diffusion length $\lambda$, found with \cref{eq:decay}, of each case is represented in the legend. 
        The inset shows the total spin accumulation and the dashed lines indicate the thermal equilibrium spin accumulation  $c$.}
        \label{fig:tempFit}
\end{figure}

\begin{figure}[htbp]
        \centering
        \includegraphics[width=\linewidth]{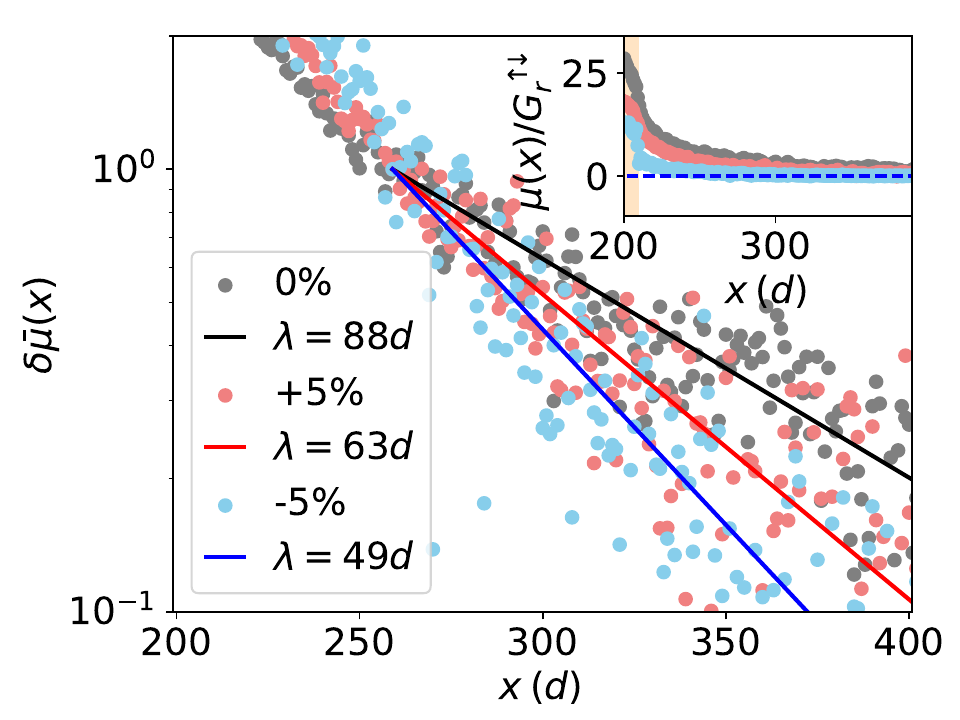}
        \caption{Spatial dependence of the electrical SHT-induced nonequilibrium spin accumulation for three magnetic states under different strain conditions. In the main plot, the y-axis is on a logarithmic scale. The extracted spin diffusion length $\lambda$, found with \cref{eq:decay}, of each case is represented in the legend.
        The inset depicts the spatial dependence of the nonequilibrium spin accumulation on the nonlogarithmic axis.}
        \label{fig:torqueFit}
\end{figure}

We utilize the material parameters listed in \cref{{tab:TransportSimParams}} to simulate magnon transport in our system with a nonlocal geometry as illustrated in \cref{fig:setupTransport}.
Figures \ref{fig:tempFit} and \ref{fig:torqueFit} depict the outcomes of atomistic spin dynamics simulations illustrating the spin signal, \cref{eq:spinAccum}, at the detector for SSE and SHT mechanisms, respectively, in three magnetic states of the CrCl$_3$ bilayer.

\emph{(i) SSE mechanism}: 
To introduce a temperature gradient across the system, we set the temperature of the injector at $T_\text{inj}=\SI{5}{K}$. 
In the biaxial EP and uniaxial OOP FM states, which are magnetic ground states under 0\% and +5\% strains, respectively, thermally excited incoherent magnons carry a net spin angular momentum that diffuses through the system and creates a spin signal at the detector. Figure \ref{fig:tempFit} shows the spatial dependence of the spin accumulation for two FM states.   
In the EP AFM state of the CrCl$_3$ bilayer, which is the magnetic ground state under a compressive strain of -5\%, two AFM magnon eigenstates in \cref{fig:dispersionrelations}(c) are roughly linearly polarized and thus in principle do not carry any spin angular momentum. 

It is worth mentioning that even in uniaxial easy-axis AFM systems with two degenerate circularly polarized magnon eigenmodes, the SSE mechanism is unable to generate a finite spin signal. This is due to the equal thermal population of the two degenerate magnon branches \cite{PhysRevB.92.180414,PhysRevB.93.014425}. 

As mentioned earlier, the thermal spin signal in SSE experiments is an even component of the spin voltage, $\delta \bar{\mu}=(\delta \bar{\mu}(x,+\bm{p})+\delta \bar{\mu}(x,-\bm{p}))/2$. Therefore, for a FM state, the background thermal spin signal, $c$, is finite in the spin signal generated by the SSE mechanism, as shown by the dashed lines in the inset of \cref{fig:tempFit}.
We extract the magnon diffusion length for the SSE mechanism by fitting the spin accumulation with \cref{eq:decay}, see \cref{tab:lambdas}. 


\emph{(ii) SHT mechanism}:
In the nonthermal SHT mechanism, most of the generated magnons have a well-defined wave vector and frequency determined by the amplitude of the spin torque and the geometry of the injector \cite{AFMtransportPRBarticle}. Hence, the magnons in the SHT mechanism are not as incoherent as in the thermal SSE mechanism.
In the SHT mechanism, the spin polarization $\bm{p}$ must be parallel to the magnetic ground state at a finite temperature background, see \cref{subsecTransportmodelling}. We set a background temperature $T_0=\SI{1}{K}$ in our simulations for the SHT mechanism. As we discussed earlier, in SHT experiments, the odd component of the spin voltage, $\delta \bar{\mu}=\delta \bar{\mu}(x,+\bm{p})-\delta \bar{\mu}(x,-\bm{p})$, is measured, and thus the background thermal spin signal does not appear in the spin accumulation data.

Figure \ref{fig:torqueFit}, shows the spatial dependence of the spin accumulation for three magnetic states. 
We find that in the SHT mechanism, we also get a finite spin signal in the biaxial EP AFM case. 
As we mentioned before, in the biaxial EP AFM case, each magnon branch is linearly polarized, which means that no spin angular momentum can be carried by them. However, in a biaxial EP AFM system, the SHT mechanism may still generate a finite spin signal through a coherent beating oscillation between two orthogonal linearly polarized magnon eigenmodes \cite{AFMtransportPRBarticle, lebrun_long-distance_2020, han_birefringence-like_2020}.
A finite spin signal can be observed if two magnon modes, with similar frequency but different wave vectors, on the two acoustic-like magnon branches, with linear polarization, pair up and combine to an effective elliptical polarized magnon mode. Therefore, the spin torque in this case must be strong enough to excite a pair of magnons at two acoustic-like magnon branches around the $\Gamma$ point. This situation is different in the other two FM cases, where it is enough to only overcome the magnon band gap of the lower acoustic-like magnon branch. 

Such finite magnon spin transport in EP AFM systems has recently been observed in the EP phase of hematite thin films \cite{lebrun_long-distance_2020, han_birefringence-like_2020}. Similar experiments in 2D EP AFM cases will shed light on the exotic magnon transport within such a system.

We extract the magnon diffusion length in three magnetic states for SHT mechanism with fitting the spin accumulation with \cref{eq:decay}, see \cref{tab:lambdas}.

\begin{table}[htbp]
\centering
\caption{The magnon spin diffusion length normalized to the lattice constant, $\lambda/d$, of three magnetic states for SSE and SHT mechanisms}
\begin{tabular}{@{}cccc@{}} \hline
		Mechanism/Strain      & \textbf{0\% } & \textbf{+5\% } & \textbf{-5\% } \\ \hline
       Thermal SSE & $68.3\pm0.1$ & $57.4\pm0.1$ & - \\
       Electrical SHT & $88\pm3$ & $63\pm3$  & $49\pm 3$ \\ 
        \end{tabular}
	\label{tab:lambdas}
\end{table}
 
Note that all transport simulations have been conducted at the same background temperature $T_0=\SI{1}{K}$, while the critical temperatures, or equivalently the magnon band gap at the $\Gamma$ point, of the three magnetic states are different; see \cref{eq:Tc}. This means that the effective temperature of magnons, or number of thermal magnons, in these three cases are different. The effect of this can be seen in \cref{fig:tempFit}, where there is a larger background spin signal, as shown by the dashed lines in the inset, in the unstrained EP FM state (gray) compared to the $+5\%$ strain with OOP FM state (red).

In general, high-frequency magnons have shorter lifetimes compared to low-energy ones \cite{chumak2019fundamentalsbook,UlrikeThesis,UlrikeSSE2}.
Our analysis shows that the magnon spin diffusion length, listed in \cref{tab:lambdas}, is shorter in the SSE mechanism compared to the SHT mechanism. This can be explained by the fact that in the thermal SSE mechanism, mostly incoherent magnons with a wide range of frequencies and momenta are generated, whereas in the electrical SHT mechanism, mainly low-frequency coherent magnon modes with a longer lifetime are generated. 

For magnons excited by the SHT mechanism, the lifetime is the largest in the unstrained case which has a lower magnon band gap and thus lower frequencies compared to magnons of the strained cases with higher magnon band gaps, see \cref{fig:dispersionrelations}.

Although the quantitative magnon spin diffusion length $\lambda$ found in this article is tied to our chosen material parameter set, we expect that the qualitative behavior holds for a large range of vdW magnetic materials.

\section{Summary and concluding remarks} \label{sec:conclusion}
In this article, we have studied the effect of strain on a CrCl$_3$ bilayer using atomistic spin dynamics simulations. First, we computed magnon dispersions and critical temperatures of the three magnetic states of the bilayer system.
Second, we computed spin signals generated by thermal SSE and electrical SHT mechanisms in these three magnetic states. 
From the results we conclude that the unstrained bilayer shows the longest propagation length and the lowest critical temperature. 

We show the high tunability of magnon dispersion relations, magnetic state, critical temperature, and spin signal using mechanical strains in bilayer vdW magnetic systems. Based on our research for CrCl$_3$ bilayer, we propose that the strain-dependent magnetic states in layered vdW magnetic systems can be monitored by spin transport measurements. 

These magnetic layers also serve as an intriguing platform for studying magnon propagation in the presence of nonuniform mechanical strains that may create the coexistence of different magnetic states in the same system. These characteristics demonstrate the significant potential of vdW magnetic layers for the next generation of magnon-based nanotechnology.

%-----------------------------
\section*{acknowledgments}
%-----------------------------
V. B. acknowledges R. F. L. Evans and A. Meo for helpful discussions. 
This work has been supported by the Norwegian Financial Mechanism 2014 - 2021 under the Polish - Norwegian Research Project NCN GRIEG “2Dtronics” no. 2019/34/H/ST3/00515.
%\newpage

\appendix

\setcounter{secnumdepth}{0}

\renewcommand{\thefigure}{A\arabic{figure}}
\setcounter{figure}{0}
\onecolumngrid

%--------------------------------
\section{APPENDIX: COMPUTING CRITICAL TEMPERATURE}
\label{sec:Tc}
\begin{figure}[h]
    \centering
    \includegraphics[width=0.5\columnwidth]{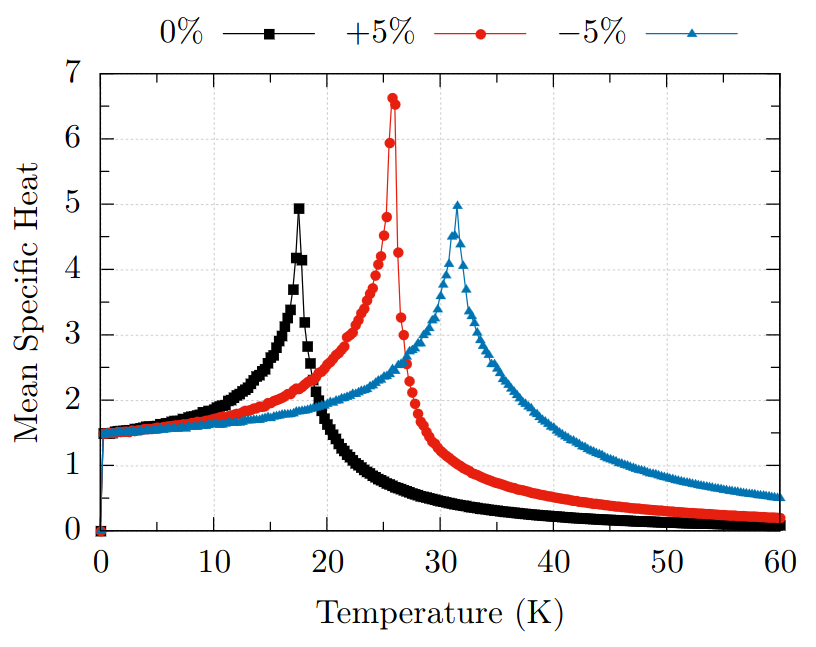}
    \caption{The mean specific heat for the three magnetic states, see \cref{tab:anisos}. We estimate the critical temperatures as $T_{\rm{c}}^\text{0\%} \approx \SI{17.8}{K}$ for the EP FM state, $T_{\rm{c}}^\text{+5\%} \approx \SI{26.0}{K}$ for the OOP FM state, and  $T_{\rm{c}}^\text{-5\%} \approx \SI{31.8}{K}$ for the EP AFM state.}
    \label{fig:Tc}
\end{figure}

%--------------------------------
The magnetic ground state for each strain case is obtained using a classical Monte Carlo algorithm with a zero-field cooling procedure \cite{vampire}.
To determine the critical temperature, $T_{\rm{c}}$, we use specific heat calculations implemented in the \vampire code.  In \cref{fig:Tc}, we present the temperature-dependent specific heat for three magnetic states. The critical temperature for each case is determined by the divergence in the corresponding specific heat. The value of the critical temperature in 2D magnetic systems is mainly governed by the magnetic anisotropies and NN Heisenberg exchange interactions. These parameters are smallest in the unstrained case and largest at $-5\%$ strain, as also reflected in their critical temperatures.

\twocolumngrid
%--------------------------------
\bibliography{literature}
%--------------------------------
\end{document}